\documentclass[aps,floatfix,nofootinbib]{revtex4}
\usepackage{latexsym}
\usepackage{mathptm}
\usepackage{amsmath}
\usepackage{amssymb}
\usepackage{graphicx}
\def\beq{\begin{equation}}
\def\eeq{\end{equation}}
\def\beqn{\begin{eqnarray}}
\def\eeqn{\end{eqnarray}}
\newcommand{\f}{\begin{equation}}
\newcommand{\ff}{\end{equation}}

\begin{document}

\thispagestyle{empty}

\title{Third road to the OPERA: a tunnel after all?}
%perspective on the OPERA anomaly\\
%{\small An onset scale and no anomalous Cherenkov-like processes
%for the deformed-Lorentz-symmetry
%perspective on the OPERA anomaly}\\
%{\small {\underline{Would}} OPERA neutrinos {\underline{require}} a preferred frame? }}
\author{Giovanni Amelino-Camelia\\
Dipartimento di Fisica, Universit\`a ``La Sapienza" and Sez.~Roma1 INFN, P.le A. Moro 2, 00185 Roma, Italy}

\begin{abstract}
The debate on OPERA has been polarized on two extremes: we should either assume ``OPERA
is wrong" or speculate that our formulation of the laws of physics needs a major overhaul.
I here argue that a more vigorous effort should also be directed toward speculating
about manifestations of laws that are already known, but whose implications are poorly
  understood in OPERA-type contexts.
My preferred example of this sort is that the OPERA result might be the
first observation of a previously unnoticed effect,
belonging to a family of effects which includes those responsible for
the well-known peculiar properties of the ``tunneling time".
Tentative support for this specific hypothesis comes from the presence of evanescence
and ``postselection", and the observation
that the fraction of neutrinos absorbed in rock on the way from CERN
to LNGS is of a few parts in $10^5$, {\it i.e.} comparable to the fraction of the overall travel
time by which OPERA neutrinos are found to reach LNGS earlier than expected.
This leads me to stress that the OPERA result is still consistent
with {\underline{sub}}luminality of a suitably ``postselection-corrected" velocity.
Adopting the popular (but here disfavored) interpretation of the OPERA result as
 a ``signal velocity"
one could obtain superluminal values of the postselection-corrected velocity
in some technologically feasible OPERA-like setups, a prediction which I propose to test.
I also propose a semi-heuristic formula for the ``rock dwell time" and find that
it gives a result in rough agreement with the OPERA travel-time measurements. This suggests that
effects similar to the ones found for OPERA's muon neutrinos
should also be found,
but with tangibly different magnitude,
for muon anti-neutrinos under similar experimental
conditions. And I argue that, whether or not they end up being
relevant for
the OPERA anomaly, the effects here considered should be significant in the
Planck-scale realm.
\end{abstract}

\maketitle

%\vskip -0.95cm

\tableofcontents

\newpage

\setcounter{page}{1}
\pagestyle{plain}

\section{A qualitative/semi-quantitative description of OPERA's travel-time measurements}
The OPERA collaboration recently reported~\cite{opera} evidence of
superluminal behavior for muon neutrinos with energies of a few
tens of GeVs: $\mathsf{v} - 1 \simeq (2.37 \pm 0.32 [stat] ~^{+0.34}_{-0.24} [sys]) \cdot 10^{-5}~~~$
 (I use units such that the speed of light is $c=1$).

It was noticed early on~\cite{whataboutopera}
that even allowing for
special-relativistic tachyons (themselves not the mildest of novelties)
one could still not provide a description of the result.
And then studies of the OPERA anomaly have been largely polarized on two extremes:
either ``OPERA
is wrong" (see, {\it e.g.},  Refs.~\cite{wrong1,wrong2,cohenglashow,gonzaOPERA,bietal,cowsiketal})
or we should speculate that our formulation of the laws of physics needs a major overhaul
(see, {\it e.g.}, Refs.~\cite{giudice,operaLED,operaELLIS,operaDSR,dimitriDSROPERA,bomaOPERA})
in an otherwise apparently familiar regime, a regime
where, {\it e.g.}, Fermi's description
of weak interactions still applies.

 I here take as working assumption that the results reported by OPERA are correct
and that not enough of our efforts
(see, however, Refs.~\cite{brusteinOPERA,ahluwaliaOPERA})
have been directed
 toward speculating
about manifestations of physical laws which are already known,
but whose implications have not yet been fully worked out.
I propose that from this perspective it might be of particular interest to consider
the role of {\underline{evanescence}} and {\underline{postselection}} in the analysis of OPERA neutrinos.

%The mains points I articulate in the next sections are summarized
%in the reminder of this section.
Since I am presently only prepared to formulate a rather vague proposal, exposing however
some sources of encouragement to pursue it, I find
appropriate to elaborate in this first section the main thesis, postponing corollary observations
 to the following sections.

\smallskip
\indent\indent{\bf peak travel time}\\
So far some aspects of the OPERA setup that are crucial from a relativistic perspective
have not been subjected to careful scrutiny. In particular,
 {\it OPERA measures the CERN-LNGS travel time
of what? information? energy? phase? what else?}\\
Since the OPERA measurement is based on recognizing and timing 
features of the intial/prepared neutrino pulse
that are reproduced in the final/observed pulse, the travel time measured by OPERA
should be describable in terms of the ``travel time of the peak of the distribution",
a time scale which is well known in the literature on signal propagation.

This is in particular the travel time typically relevant for the description of
measurements whose focus is on
the most likely detection time of a particle
and the most likely emission time of ``that same" particle. And (as illustrated in Fig.~1)
 the ``travel time of the peak of the distribution"
 gives results in agreement with several other
 travel-time determinations
based on features of emission and detection probability distributions.

\begin{figure}[htbp!]
\includegraphics[width=0.55 \columnwidth]{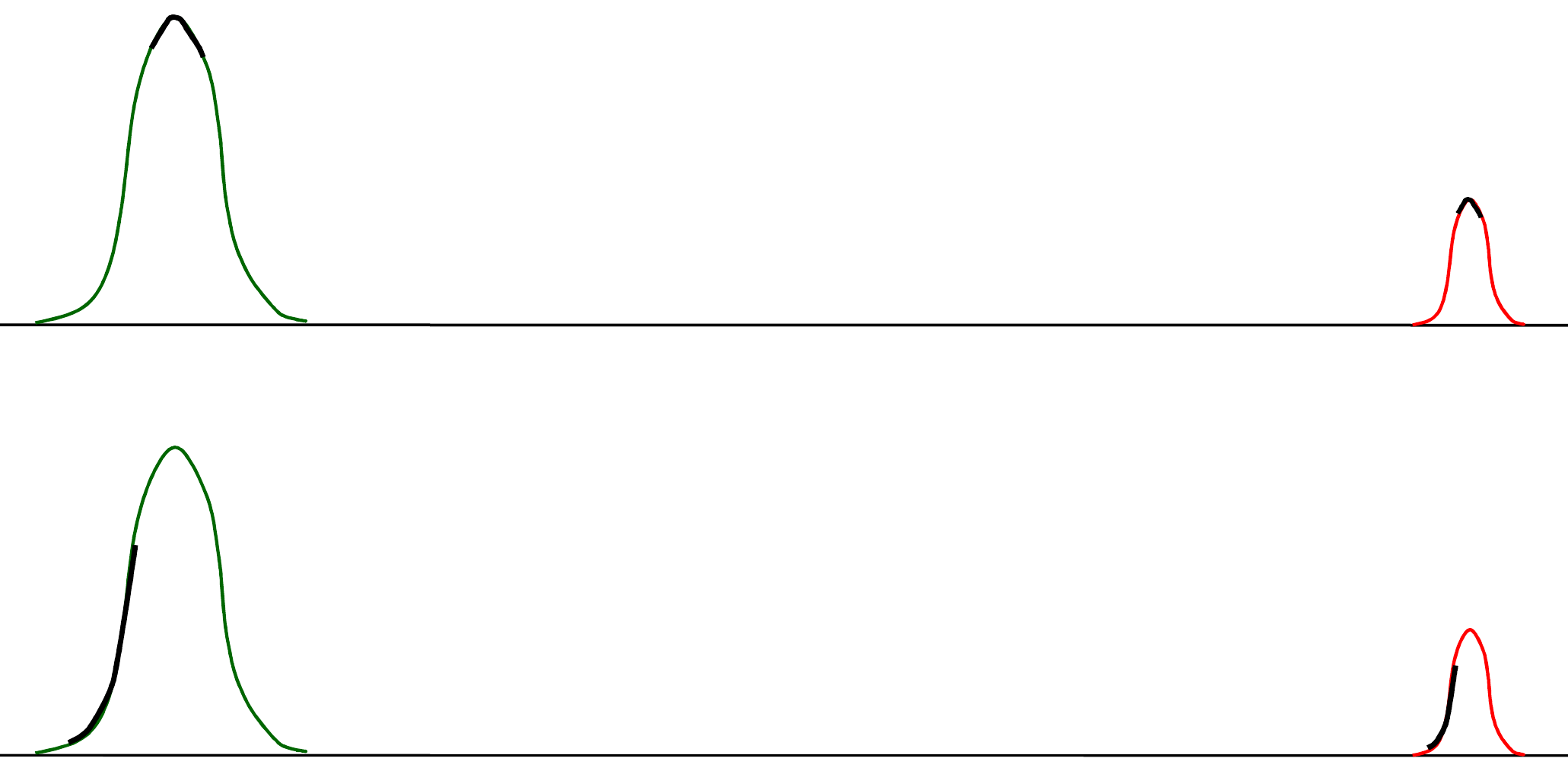}
\caption{A ``cartoonist impression" of the notion of ``velocity of the peak of the distribution".
In figure greentime is for time of emission while redtime is for time of detection.
Though the velocity of the peak of the distribution in general does not coincide with the signal velocity,
it is a very valuable observable. In particular (top panel) velocities measured in terms
of
the most likely detection time of a particle and the most likely emission time of ``that same" particle
end up being determinations of the velocity of the peak of the distribution.
And also when travel times are measured through other features, such as
 timing the edges of the emission and of the
detection distributions (bottom panel), the results usually still agree with
the ``travel time of the peak of the distribution".}
\end{figure}

\smallskip
\indent\indent{\bf not necessarily ``information travel time"}\\
The most fruitful notion of travel time of interest from a relativistic perspective
is the notion of ``travel time of information" (or ``signal travel time"),
the time it takes for information
coded through some process to be received through some other process.
 Not withstanding the fact that
information travel time is still only poorly defined from a relativist perspective~\cite{gacinprep},
if there were truly superluminal information-velocity measurement results we should go beyond special relativity.\\
But it has not been established
that OPERA measures an information travel time. As I just stressed OPERA effectively measures a
pulse-peak travel time, a ``time interval between the peaks of two different but related pulses",
and it is well known~\cite{stein2007,physrepREVIEW,recamiREVIEW} that
these pulse-peak travel times do not in general coincide
with the information travel times. 
It is also well established~\cite{stein2007,physrepREVIEW,recamiREVIEW}
that superluminal pulse-peak velocities
 can be produced by theories governed by standard Poincar\'e invariance.\\
With the facts so far available, even assuming the OPERA time measurement is accurate
and free from unknown systematic errors, special relativity is still not objectively threatened.

\smallskip
\indent\indent{\bf A case of ``generalized tunneling time"?}\\
Among the known instances in which a peak velocity was found to be superluminal,
within a perfectly Poincar\'e-invariant description, and while the information velocity
was subluminal,
particularly interesting examples are provided by studies
of ``tunneling 
time"~\cite{stein2007,physrepREVIEW,recamiREVIEW,landauerNATURE,steinbergPRL,abruptNATURE,diractunnel1,diractunnel2}
and of anomalous dispersion~\cite{anomdispPRL,anomdispNATURE}
(but there are several others).
The possibility that the OPERA result might be an unforeseen manifestation
of an effect of similar type should evidently be investigated throughly
before any claim is made about a disagreement between special relativity
and the OPERA result. And I feel that a speculative/exploratory attitude
for such studies is well justified considering that the alternative of
revising special relativity in a regime accessible to this generation of 
accelerators requires adopting a strongly speculative attitude.

From this perspective many options could be considered (some of which
are mentioned here below),
but my preferred example of this sort is that the OPERA result might be the
first observation of a previously unnoticed effect,
belonging to a family of effects which includes those responsible for
the well-known peculiar properties of the ``tunneling time".
This would locate the solution of the OPERA puzzle at what one may label as
the slippery side of the interplay between special relativity and quantum mechanics,
an arena which our current theories in principle allow us to master fully,
but for which several of the effects that could be produced have still not been recognized.
It is important to discover  experimentally
these new (previously unknown) effects  and to recognize them phenomenologically,
both because of their intrinsic interest and because of their possible usefulness
in contributing to a better understanding
of the interplay between special relativity and quantum mechanics.

The example of studies of the ``tunneling time", which shall inspire a large fraction
of my speculations, provides a crisp illustration
of how issues can get slippery in some areas of the interplay between special relativity
and quantum mechanics.\\
How much time does it take for a particle to tunnel through a barrier?\\
Of course this sort of classical-limit-inspired questions are not fully meaningful
in cases, such a quantum tunneling, where the classical limit does not exist.
But we learned a lot about the interplay between special relativity
and quantum mechanics by looking for travel times which would make
sense in the case of quantum tunneling.

In particular, for quantum tunneling there is a notion of ``velocity of the peak"
which is to a large extent analogous to the ``velocity of the peak" I mentioned 
above for OPERA neutrinos. As described qualitatively in Fig.~1,
the velocity of the peak can be well defined in appropriate studies of quantum tunneling
and for suitable experimental setups
 it is found to be superluminal, both experimentally and within descriptions of tunneling
based on theories that are perfectly Poincar\'e 
invariant~\cite{stein2007,physrepREVIEW,recamiREVIEW,landauerNATURE,steinbergPRL,abruptNATURE,diractunnel1,diractunnel2}.
While the naive description of these facts about the velocity of the peak can lead
to speculations about violations of all sorts of laws, it is not difficult to
see that in such contexts the velocity of the peak is not the ``signal velocity" (``velocity
of information") and is therefore of very limited relativistic interest.

\begin{figure}[htbp!]
\includegraphics[width=0.55 \columnwidth]{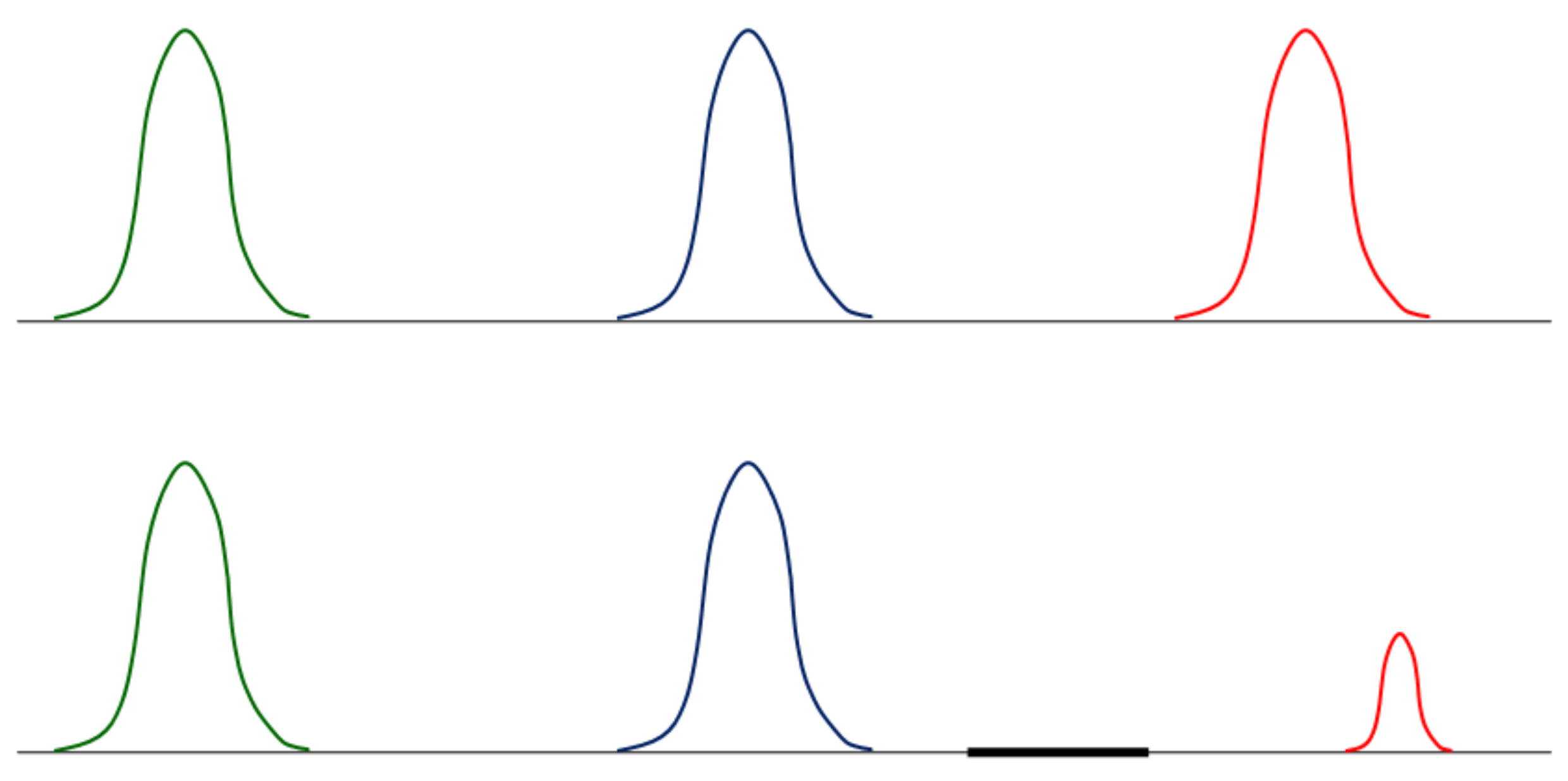}
\caption{A ``cartoonist impression" of an aspect of the tunneling-time issue, as viewed
from the naive ``peak velocity" perspective.
The figure shows a time sequence (greentime, bluetime, redtime) of descriptions
of a probability distribution for the position of a particle along a certain
spatial direction (the horizontal direction in figure). Two time sequences are
shown: the one in the top panel is for propagation without a barrier while for the
one in the bottom panel a barrier is located (but not shown in figure)
where the horizontal line is thick.
In the bottom panel at time redtime I am evidently only showing the small
transmitted peak, since also showing the large
reflected peak would affect the visibility of the figure.
This qualitative picture of ``peak velocity"
for quantum tunneling is confirmed by a
 large collection of measurement results, including some studies where interference between a 
 path without a barrier and
a path with a barrier is measured (a striking single-photon interference study is reported
in Ref.~\cite{steinbergPRL}). And these measurement results are correctly reproduced by
theoretical pictures that are perfectly Poincar\'e invariant.
An anomaly appears to be present if one focuses
on the time interval found timing the peak ({\underline{of the entire distribution}})
before reaching the barrier and timing the peak
({\underline{of the post-selected, transmitted, distribution}})
after tunneling through the barrier. Naively this would suggest
that the  particles
that do tunnel through take a time for traversing the tunnel which is shorter than the
time needed by the same particles to traverse a distance in vacuum equal to the
length of the barrier.}
%\label{RIFARELALABEL}
\end{figure}

\newpage

\smallskip
\indent\indent{\bf signal velocity versus ``velocity of the peak through a tunnel"}\\
What is the meaning of the velocity of the peak in quantum tunneling?\\
Naively one would like to apply classical-limit logics to the
quantum phenomenon at hand:
at first there is a (big) pulse
approaching the barrier and then a (much smaller) pulse is observed on the other
side of the barrier, so it {\underline{appears}} natural
to  describe the outgoing pulse as ``a part of the incoming pulse,
after traveling through the barrier". Of course in quantum mechanics (and therefore
in Nature)
there is no such notion as ``the incoming pulse
after traveling through the barrier": the only scientific notions quantum mechanics
afford us concern the (preparation of the) initial pulse
and the (measurement of the) final pulse. 

In the case of quantum tunneling,
at a more careful level of analysis these facts emerge very clearly
since the outgoing pulse is so very different from the incoming pulse, but our intuition
is nonetheless challenged by the fact that under appropriate
conditions~\cite{stein2007,physrepREVIEW,recamiREVIEW,steinbergPRL}
the distribution observed after the tunnel is (attenuated but) undistorted
with respect to the incoming distribution.
Nonetheless the barrier acts on the pulse very virulently,
with processes governing the reorganization of the incoming
distribution into two different distributions, one (strongly attenuated) ``transmitted" on the
other side of the tunnel,  and one (containing most of
 the signal) ``reflected" back toward the direction from which
the original distribution was coming from.

These observations expose the fact that the ``velocity of the pulse peak" cannot
be automatically interpreted as a signal velocity, and in particular superluminal values
of the velocity of the pulse peak cannot be seen as automatically in conflict with special relativity.
But, while it is clear that they are not automatically in conflict with special relativity,
can one show that superluminal velocities of the pulse peak are consistent with
special relativity?\\
Many argument suggest that they are~\cite{stein2007,physrepREVIEW,recamiREVIEW,steinbergPRL}.
Let me just mention one of these observations, emphasized, {\it e.g.}, in Ref.~\cite{stein2007}.
Even though ``the peak appears to arrive too early" one can show that
the results on tunneling that have been obtained experimentally
can all be reproduced in terms of a description, after exiting the barrier,
which is based on a ``response function" that fully obeys to
special-relativistic causality~\cite{stein2007}:
$$
\psi(x_f , t_f) = \int_0^\infty d\tau
 ~f(\tau)~\psi(x_i,t_f-\tau-(x_f-x_i)/c)$$
The behavior of $\psi(x,t)$ at $x=x_i$ for times $t > t_f - (x_f-x_i)/c$ does not affect in any way
the behavior of $\psi(x_f , t_f)$.\\
Evidently the peak of the distribution is not necessarily where the ``information content"
resides. On the contrary for smooth, frequency-band limited, distributions
the ``precursor tail" of the distribution
allows to infer by analytic continuation the structure of the peak:
under such conditions, even for
free propagation, by the time the peak reaches a detector it carries no ``new information"~\cite{stein2007}
 with respect to the information already contained in the precursor tail.
An example of ``new information" is present in
modified pulses containing ``abrupt" signals~\cite{stein2007,abruptNATURE},
and indeed experimentally one finds that when these new-information features are sufficiently
sharp they never propagate superluminally~\cite{abruptNATURE}
(also see Ref.~\cite{energyflowSCIENCE}).

However, one cannot stress enough that the ``velocity of the peak"
is a very useful observable, which can be robustly measured under rather conventional
experimental conditions: it is the most natural observable one would consider
when studying the tunneling time, and indeed it was measured on countless occasions
in quantum-tunneling experiments~\cite{stein2007,physrepREVIEW,recamiREVIEW,steinbergPRL}.
And, as I stressed above, it is the observable that, without even apparently planning to do so,
OPERA ends up measuring for the CNGS neutrinos.
In the relevant situations
(such as in presence of a barrier, with emitter on one side and detector on the other side),
the ``travel time of the peak" is the time interval typically found, for a given prepared distribution,
between the most likely time when (and position where) a particle  is prepared
and the most likely time when (and position where) correspondingly a particle
is detected, but nonetheless, as stressed above,
 it is not the travel time of information.

\smallskip
\indent\indent{\bf OPERA: no tunnel but evanescence and postselection}\\
Evidently my plan is to find analogies between the tunneling-time problem
and  the ``OPERA-travel-time puzzle",
but there is an evident limitation
for taking this line much further: $~~$ {\underline{where is the tunnel?}}\\
A prerequisite for a tunnel is the presence of a barrier, but in our current
%(of course to some extent heuristic)
conceptualization of neutrino propagation from CERN to LNGS there is
no barrier.
%\footnote{Claims~\cite{gelmini1,gelmini2} that OPERA neutrinos do travel from CERN to LNGS
%through an ordinary tunnel are inaccurate~\cite{opera}.}.
So ultimately this is not going to be a tunneling-time story.
As mentioned, I am conjecturing that the OPERA puzzle could be a
manifestation of a previously unnoticed
large class
of travel-time effects, (different from but) sharing some properties of
the ``tunneling time".

And, besides the nature of the travel-time measured by OPERA (which should be describable
 computing the ``pulse-peak travel time")
 there are at least two other features of the OPERA setup that could place the associated
 effect within a larger
 category of effects that includes quantum tunneling: evanescence and postselection.\\
In quantum tunneling evanescence arises
 because the probability distribution ``transits through" a classically forbidden region,
 while in OPERA some evanescence comes into play because of the fact that a fraction of
 neutrinos is absorbed in rock.
 Postselection is the other side of this story: at LNGS the OPERA detectors can only
 detect neutrinos which have not been absorbed in rock, just like the transmitted
 particles in quantum tunneling are only some of the prepared particles,
 the others being reflected and not being considered in the determination of
 the tunneling time.

\smallskip
\indent\indent{\bf   evanescence
and postselection could be non-negligible for OPERA neutrinos}\\
But evanescence and postselection are minute for the OPERA set up, how could they matter?\\
Chances are it will eventually turn out that they do not matter, but they could. With the limited
information we presently have we must consider the possibility that they could matter.\\
At present we must adopt this attitude because indeed evanescence and postselection are minute in
the OPERA setup
but also the OPERA anomaly is minute: an effect of a few parts in $10^5$.\\
And it happens to be the case that the fraction of $\sim 20 GeV$ neutrinos
absorbed in rock on the way from CERN
to LNGS, here denoted as $\xi$, is of the same order of magnitude
as the fraction of the overall travel time
by which OPERA neutrinos are found to arrive "early":
$$\xi_{(20GeV)} \equiv \Big[1 - {\cal P}_{CERN \rightarrow CNGS} \Big]_{\sim 20GeV} \simeq
~ L ~(\rho N_A) ~\sigma_{(20GeV)} ~ \simeq ~ (7 \cdot 10^7cm) \frac{2 \cdot 10^{24}}{cm^3}
 (1.4 \cdot 10^{-37}cm^2) ~ \simeq ~ 2 \cdot 10^{-5}~.$$

This ``numerological coincidence"
is very striking: of course it may be one more instance when numbers show us their
peculiar taste for jokes, but I feel we cannot look away\footnote{I am here
factoring in also the instances in the history of physics when a ``wrong hint",
an observation which should not have been viewed as a hint, accidentally pointed us
in the right direction.} at such a candidate
phenomenological hint
for a solution of the OPERA puzzle. And in any case adopting a genuine exploratory attitude
toward the OPERA anomaly we {\underline{must notice that absorption in rock
is not a small effect, not for the OPERA-neutrino result}}: the result shows an anomaly at
the level of a few parts in $10^5$ so we cannot automatically assume that another
effect of comparable significance, evanescence due to neutrino absorption in rock,
which is also present at the level of a few parts in $10^5$, could be safely neglected.

And the same reasoning applies to postselection, which in the OPERA setup is directly connected
with evanescence: the ensemble of CNGS neutrinos studied by OPERA is postselected in the sense
that only those who manage not to be absorbed in rock can partecipate in the travel-time
measurement. It is a case of minute postselection: nearly all CNGS neutrinos have a chance
of contributing to the OPERA measurement, with only a few every $10^5$ (according to my estimate)
being precluded from contributing. Again, since the effect reported by OPERA is at the
level of a few parts in $10^5$, one cannot assume without question that
 this minute role for postselection should be negligible.\\
 And conceptually it is interesting to contemplate this issue: usually postselection
 attracts attention in the physics literature when it is sizable, and ``weak measurements"~\cite{weakmeas1,weakmeas2}
 are involved, whereas here one could contemplate a role for a minute amount of postselection,
 probably describable in terms of some sort of ``quasi-strong measurement".
It is known that ``weak measurements" can produce ``large anomalies"~\cite{weakmeas1,weakmeas2}
for some observables. The type of ``quasi-strong measurements" which I am considering
will probably instead only produce small anomalies. And this hypothesis is a good
match for the OPERA puzzle, where
the anomaly is striking conceptually but quantitatively small.

\smallskip
\indent\indent{\bf postselection-corrected speed of OPERA neutrinos is not superluminal}\\
Having stressed the role of (minute but nonnegligible) postselection in OPERA,
it is then natural to wonder what would be the average velocity found by OPERA if it
was not postselecting. In first-quantized quantum mechanics (and experiments
not involving particle production) it makes sense to contemplate how the average position
of the particle $<x>$ evolves with time, especially when the velocity of the peak is found
to be ``superluminal" (for simplicity I am assuming here only one spatial direction, $x$).
Through case-by-case analysis one finds that
(see, {\it e.g.}, Ref.~\cite{stein2007})
 the average speed obtained from $<x>$ over the whole ensemble (no postselection)
never gets superluminal, even in cases where
the ``velocity of a peak" is found to be superluminal.
This is in particular automatically ensured for studies of the particles
transmitted and reflected by a barrier, since most of the particles in the initial distribution
are reflected, so
even if the very few that are transmitted have a formally superluminal velocity (of the peak)
the contributions to $<x>$ from reflected particles dominate and the net result
is indeed such that
the average speed obtained from $<x>$
never gets superluminal.

In order to generalize this notion to an OPERA-type setup one must find a way to handle
the fact that, as a result of neutrino absorption in rock,
in the OPERA setup the number of particles is not conserved.
Let me propose a way to adapt this notion to OPERA-like contexts within
a formulation that involves directly
the notation $\xi$ which I already introduced above\\
\indent $\xi \equiv$ fraction of CNGS neutrinos
absorbed in rock on the way from CERN
to LNGS\\
and denoting by $\delta$ the OPERA anomaly\\
\indent $\delta \equiv \mathsf{v} - 1$, where $\mathsf{v}$ is the neutrino ``velocity"
 measured
by OPERA.\\
I gave a rough estimate above of $\xi$, which is $\xi \sim 2 \cdot 10^{-5}$,
and as mentioned the result reported by OPERA is
$\delta \simeq (2.37 \pm 0.32 [stat] ~^{+0.34}_{-0.24} [sys]) \cdot 10^{-5}$. So
it should be noticed that, even taking the OPERA result at face value,
the data are still consistent with the possibility
$$\xi \geq \delta$$
Now let us compute $<x>$ adopting the following heuristic formula
$$<x>_{OPERA} \simeq (1-\xi)(1 + \delta) t $$
essentially counting as contributions to $<x>_{OPERA}$ only the ones from neutrinos
that would have avoided absorption in rock, reaching (but of course not necessarily being detected)
LNGS, and this gives the $(1-\xi)$ factor, but allowing for them a ``speed" as reported
by OPERA, and this gives the $(1 + \delta)$ factor.\\
One then finds, since both $\xi$ and $\delta$ are small,
$$<x>_{OPERA} \simeq t +(\delta-\xi) t~.$$
So, as long as data are consistent (within errors) with $\xi \geq \delta$,
the average speed obtained from OPERA data with
my prescription for a ``postselection-corrected $<x>$"
is {\underline{not superluminal}}, even though $\delta > 0$.\\
Looking also beyond OPERA, to include other setups that might benefit from this type
of reasoning, I am led by this argument to propose that there should be some interest
into the following related definition of ``postselection-corrected velocity":
$$\mathsf{v}_{corr} = {\cal P}_{select} \mathsf{v}_{select}$$
where ${\cal P}_{select}$ is the probability of a particle in the ensemble of
being ``selected" for the measurement of $\mathsf{v}_{select}$ (probability of tunneling,
for tunneling particles, probability of avoiding absorption in rock, for OPERA neutrinos, and so on)
and $\mathsf{v}_{select}$ is the measured ``velocity of the pulse peak".\\
To my knowledge such a notion of ``postselection-corrected velocity" had not been used
 or proposed in the previous literature, but I find that at least for the
 most studied cases of ``superluminal peak velocity" one would still determine
 subluminal values for  my ``postselection-corrected (peak) velocity".
And the simple equations for $<x>_{OPERA}$ which I showed above indicate that
even the OPERA result is still consistent with a subluminal
postselection-corrected velocity.\\
What is perhaps intriguing in these observations is that it seems no general theorem 
could ensure that $\mathsf{v}_{corr}$ should always take subluminal values.
Similar theorems are natural for signal velocity, because of its physical
meaning from a relativistic-causality perspective, but I cannot imagine
analogous theorems to apply to my postselection-corrected (peak) velocity since in
principle it may well be superluminal without violating relativistic causality.
The fact that experimentally it has always taken subluminal values
therefore raises an interesting challenge either for theoretical understanding
or for experimental techniques. I shall return on this challenge in later parts of this manuscript.

\smallskip
\indent\indent{\bf a semi-heuristic estimate of the ``rock dwell time" matches the OPERA result}\\
I have so far provided indirect hints that the OPERA travel-time result 
might admit a description and interpretation
 that are very different
from the ones most commonly adopted in the recent OPERA-inspired literature. And I argued
that the associated challenges of formalization and interpretation
are to some extent analogous to challenges familiar in various contexts, including
studies of the tunneling time. I have postponed the main challenge: to fully support
this interpretation one should compute within relativistic quantum field theory the travel time measured
by OPERA and find a result in agreement with the experimental result. In parts of the
following sections I shall consider the challenges and the opportunities for performing
such a calculation. Here I want to report an observation concerning a ``rock dwell time"
which is not exactly the travel time measured by OPERA, but may nonetheless
be relevant for a preliminary analysis.

In studies of quantum tunneling it is very common to refer to the ``barrier dwell time".
This is a time scale quantum mechanics allows us to compute, when the time taken to
travel through a certain region is of interest. Considering the region of interest is along
the $x$ axis, between $x=0$ and $x=L$, the dwell time then is
(see, {\it e.g.}, Ref.~\cite{stein2007} and references therein):
$$t_d \equiv \int^{\infty}_{-\infty} dt \int^{L}_{0} dx~ |\psi(x,t)|^2$$
where
$$\int^{L}_{0} dx~|\psi(x,t)|^2$$
is the instantaneous probability that the particle ``is inside the region",
and then $t_d$ is a candidate for the ``expected value of the time spent by the particle
inside the region".

In the classical limit this observable gives the time deterministically needed
for the particle to travel through the region (for example for a nonrelativistic
particle one finds that in the classical limit $t_d = m L/p$, where $p$ is the spatial
momentum of the particle and $m$ its mass). Within quantum mechanics, particularly in cases
(such as quantum tunneling)
where the classical limit is problematic, the dwell time is
 not a fully satisfactory characterization of
the travel time~\cite{stein2007}. For example, the barrier dwell time
provides an unsatisfactory notion of tunneling time,   because it includes
 at once contributions from both the case of
 transmitted particle and the case of reflected particle (whereas
 one would like a notion
 of  tunneling time receiving no contribution from reflected particles).

 It is nonetheless an established~\cite{stein2007,physrepREVIEW,recamiREVIEW}
 experimental fact, confirmed by model calculations,
that in several (though not all) different experimental setups
the dwell time gives a reasonably reliable rough 
approximation of the ``travel time of the peak through the region",
which is the travel time here of interest.
So I shall speculate that some appropriate definition of ``rock dwell time" should
give a rough description of the OPERA results, just like for studies
of quantum tunneling the ``barrier dwell time"
often gives a rough description of the ``travel time of the peak through the tunnel".
Note however that
I am assuming OPERA measures the travel time of the pulse
peak from CERN to LNGS, but evidently
there is no standard barrier along the way. Moreover, just because of absorption in rock,
 the effect which is the main focus of this manuscript,
 the OPERA setup is such that the number of particles is not conserved:
 in tunneling the transmitted beam is (very severely) attenuated while conserving the
 total number of particles (the missing particles are reflected), instead in the OPERA setup
 the beam that reaches LNGS is (minutely) attenuated because of absorption.
Taking as working assumption that these differences should not prove too costly,
I can make a tentative estimate of
 the "rock dwell time" for OPERA neutrinos:
\begin{equation}
t_{d,OPERA} \simeq \int^{\infty}_{-\infty} dt \int^{L}_{0} dx~ |\psi_{OPERA}(x,t)|^2
 \sim \int^{L}_{0} dx~e^{-\xi_\partial x} \simeq  L -\xi_\partial L^2/2
 \simeq \left(1-\frac{\xi}{2}\right) L
 \label{dwellOPERA}
\end{equation}
 where $e^{-\xi_\partial x}$ describes the absorption, $\xi$ is the feature I already
 considered a few times above, also estimating it to be $\xi \sim 2 \cdot 10^{-5}$
(so I estimated that absorption of muon neutrinos of $~20 GeV$ in rock
has $\xi_\partial \simeq \xi/L \sim 2 \cdot 10^{-5}/(732Km)$).

On the basis of (\ref{dwellOPERA}) one would conclude that the ``rock dwell velocity"
of OPERA neutrinos is
$$v_d\sim 1 + \xi_\partial L/2 \sim 1 + \xi/2 \sim 1 + 10^{-5}$$
So by this procedure one gets a ``rock dwell time" which is in rough agreement
with the travel time measured by OPERA. This
rough agreement is rather striking, considering the many simplifying approximations
and assumptions I made, and also considering that in general the agreement between dwell time
and peak-travel time  is (when at all present) only approximate,

\smallskip
\indent\indent{\bf phenomenology}\\
My proposal of analyzing the OPERA result from the viewpoint of an example of travel-time
issue, as for the tunneling time and other known peculiar travel-time effects, and particularly in relation to
evanescence and postselection, remains in this manuscript at a poorly formalized level.
Because of some known limitations in our mastery of 
relativistic quantum field theory
 a rigorous analysis adopting my perspective
 appears to be very challenging. One would
need to do an analysis of ``travel time of the peak of the distribution"
within relativistic quantum field theory, taking into account also the many-body nature of
the OPERA setup (many neutrinos in each bunch sent out from CERN, many target nucleons
along the way). Perhaps this can be done in the not-so-distant future. I have of course
efforts in that direction ongoing, and hopefully this manuscript will motivate others to try.
It appeared however appropriate to publicize this tentative picture of the OPERA anomaly because,
as I shall now emphasize, it can be interestingly investigated experimentally, even without
a fully worked out model.

As experiments for investigating the OPERA anomaly are getting set up it appeared
there would be some purpose for illustrating my argument, even at the present preliminary
stage of development, indeed because I find that, in spite of the preliminary status,
it raises some scenarios and some challenges which 
might perhaps already be intriguing enough to pursue them experimentally.
And, now that experiments
following up on OPERA are being set up, it is the right time for discussing the many
different types of experimental studies which might shed light on OPERA.

\smallskip
\indent\indent{\bf absorption and dwell-time phenomenology: testing OPERA with antineutrinos}\\
Since evanescence and absorption are  key ingredients of my argument,
any experiment recreating roughly the same conditions as OPERA but with different
strength of absorption could shed light on the possible relevance of my thesis.
Many ways to do this are obvious. For example, if (prematurely) flirting with
large-budget ideas, one could contemplate travel-time studies of muon neutrinos traveling through
regions of higher density of nucleons, such as the Earth core.
Another example is the one
of OPERA-like setups for neutrinos of higher energies,
so that the implications of the corresponding increase of the neutrino-nucleon
cross section (see, {\it e.g.}, Refs.~\cite{liparilusignoli,pdg}) could be
used to test the possibility that indeed absorption is playing a role.

Among these many possibilities I believe there is one that deserves particular interest,
and takes relatively little effort:
 one could run an experiment completely analogous to OPERA with
muon antineutrinos rather than muon neutrinos. The advantage of this is that
in a large majority of even the most speculative new-physics models neutrinos and antineutrinos
would have the same OPERA-relevant properties. So many new physics interpretations predict that
a ``OPERA for antineutrinos" should find the same anomaly that the
original ``OPERA for neutrinos" reported in Ref.~\cite{opera}.\\
On the contrary in the proposal I am here putting forward these measurements
of time delays are interpreted in connection with absorption (and the associated features
of evanescence and postselection), and
accordingly one would expect a ``OPERA for antineutrinos"
to find results tangibly different from the ones of the
original ``OPERA for neutrinos":
at energy scales comparable to the ones of OPERA the antineutrino-nucleon
cross section is significantly different from the
neutrino-nucleon
cross section (see, {\it e.g.}, Ref.~\cite{liparilusignoli,pdg})
$$ \sigma_{\,_{{\overline{\nu}}_\mu}} \simeq \frac{1}{2} \sigma_{\,_{\nu_\mu}}$$
Even without a reliable derivation of the effects of evanescence/absorption and postselection
on OPERA-type setups, according to my perspective it would be inevitable for
 a change of roughly a factor 2 in cross section
to have interesting consequences: the ``anomalous" effect should not be eliminated
but  its magnitude should change significantly.

If one could trust my rough comments on the ``rock dwell time", and if we could
assume its relevance
for the determination of the ``pulse-peak travel time" measured by OPERA,
one would then expect that an experiment exactly like OPERA but replacing muon neutrinos
with muon antineutrinos should find an effect which is roughly one half of
the effect reported by OPERA for neutrinos.
But evidently my ``rock dwell time" can at best be a rough estimator of OPERA-like travel times,
so it would be unwise to trust it that far. It may well be that for antineutrinos one would
find a stronger rather than smaller effect. But still it should be unavoidable to find
the same effect also for antineutrinos, and it should be unavoidable that
the magnitude of the effect is tangibly different. These (however qualitative)
expectations differentiate my proposal very strongly from other
proposed ``solutions of OPERA", according to which  ``OPERA for antineutrinos"
should  find results identical to the ones of the
original ``OPERA for neutrinos".

\smallskip
\indent\indent{\bf phenomenology challenging the subluminality of the postselection-corrected velocity}\\
Another potentially intriguing possibility for phenomenology comes from my notion
of ``postselection-corrected subluminal-velocity bound".
As stressed above I do not expect there could be a theorem ensuring that
my notion of postselection-corrected velocity should always be subluminal.
But it is subluminal even in many cases where the ``pulse-peak velocity" is superluminal.
 So it may well remain subluminal in all
 OPERA-like setups. And if it does, if we gather experimental
information in support of this postselection-corrected subluminality bound, then in turn
we would have strong encouragement for assuming that somehow
evanescence/absorption and postselection are playing a role in the OPERA anomaly.

The starting point for possible future tests of this conjectured bound
is the fact that I surprisingly found  OPERA neutrinos
to be very close to ``luminal" postselection-corrected
velocity. In general my definitions and arguments imply that
\begin{eqnarray}
&& \xi > \delta ~~~ \Longrightarrow ~~~ \mathsf{v}_{corr} <1 ~~~\textrm{(subluminal)}
\nonumber\\
&&\xi = \delta ~~~ \Longrightarrow ~~~ \mathsf{v}_{corr} =1 ~~~\textrm{(luminal)}
\nonumber\\
&&\xi < \delta ~~~ \Longrightarrow ~~~ \mathsf{v}_{corr} >1 ~~~\textrm{(superluminal)}
\nonumber
\end{eqnarray}
where I denoted again with $\mathsf{v}_{corr}$
the postselection-corrected velocity I defined earlier.\\
What I surprisingly found is that the OPERA result is in some sort of ``limbo" as far
as determining the nature of my postselection corrected velocity:
the OPERA result is still consistent within errors both with $\mathsf{v}_{corr} \leq 1$
and with $\mathsf{v}_{corr}> 1$.

It is important  to notice that according to
 previously attempted descriptions of the OPERA anomaly this
balance $\xi \simeq \delta$ would have to be
viewed as accidental. Instead it fits
rather naturally within my proposed interpretation/description of the OPERA anomaly:
my argument essentially establishes a relationship between $\xi$ and $\delta$ (which are
completely unrelated in all other proposed solution of the OPERA anomaly), so, even without
a result establishing the form of this relationship, the fact that $\xi \simeq \delta$
in the OPERA case would not be surprising.

Evidently one way to discriminate between my proposal and other proposed solutions of the
OPERA anomaly is to collect data in experimental setups which are similar to (and yet tangibly
different from) the OPERA setup:\\
$\star$ according to the popular (but here disfavored) interpretation of the OPERA results
in terms of a ``physical particle velocity" $\delta$ should be independent of $\xi$, whereas\\
$\star$ according to my proposal different experiments attempting to measure the ``velocity"
of neutrinos and antineutrinos should find different values of $\delta$, gradually establishing
the form of the relationship between $\delta$ and $\xi$ (which I am presently unprepared to predict).

It would be particularly
interesting from this perspective to obtain data from a ``shorter OPERA": taking as starting point
the OPERA neutrino-velocity setup one would change only the distance between emitter and detector,
keeping all other aspects unchanged (muon neutrinos in the same energy range, same
density of the rock...).
Since OPERA involves a distance of $732 Km$ it would be particularly interesting to
look at shorter distances, in the range of, say, $250$ to $300Km$.
In such cases the characteristic $\xi$ (fraction of particles which is absorbed)
would clearly be smaller, $\xi < \xi_{OPERA}$,
but what would be found for $\delta$?\\
If $\delta$ corresponds to a ``physical particle velocity" all such experiments
should find values of $\delta$ consistent with the OPERA determination, and this could be tested
with a ``shorter OPERA" with a
distance from emitter to detector of, say, $2/5$ of $732Km$ ($\sim 295 Km$):
$$\xi_{2/5OPERA} \simeq \frac{2}{5}\xi_{OPERA}~,~~~\delta_{2/5OPERA} \simeq \delta_{OPERA}
~~~\Longrightarrow ~~~\xi_{2/5OPERA}-\delta_{2/5OPERA} \simeq - \frac{3}{5} \delta_{OPERA} < 0$$
where I used the fact that, as stressed above, it happens to be the case
that $\xi_{OPERA} \simeq \delta_{OPERA}$, and I also used the fact that by
definition $\xi_{2/5OPERA} \simeq (2/5)\xi_{OPERA}$ together with
the physical-particle-velocity assumption $\delta_{2/5OPERA} \simeq \delta_{OPERA}$.\\
Two points must be stressed here:\\
$\star$ indeed it would not take much to establish that $\xi_{OPERA} \simeq \delta_{OPERA}$ is
accidental, since for an emitter-detector distance half as big or twice as big
as OPERA's emitter-detector distance one would then find $\xi - \delta$ taking values
 tangibly different from zero 
(if the energy range of the neutrinos is similar to OPERA's and
assuming the interpretation
in terms of a ``physical particle velocity" is correct);\\
$\star$ in particular if the emitter-detector distance is significantly {\underline{shorter}}
than OPERA's emitter-detector distance the hypothesis
that $\delta$ is a manifestation of a ``physical particle velocity" would produce
 a violation of my conjectured subluminality bound for the postselection-corrected velocity
 (this is why I highlighted the fact that $\xi_{2/5OPERA}-\delta_{2/5OPERA} < 0$ in that hypothesis).

So we do have, in perfectly doable experiments, ways to distinguish between
the more popular physical-particle-velocity description of the OPERA result and
my preferred interpretation linked to absorption and postselection.
Moreover, the physical-particle-velocity description of the OPERA result {\underline{predicts}} that
there will be instances of measured superluminal values
for my notion
of postselection-corrected velocity. But in order to expose such
 superluminal values
for the postselection-corrected velocity one should rely on
emitter-detector distances shorter than OPERA's. This is amusing since
the objective of achieving higher accuracy
for OPERA-type velocity measurements would lead one to favor longer
emitter-detector distances, but it is easy to see that with longer
emitter-detector distances the subluminality of
my postselection-corrected velocity would not be challenged.

\smallskip
\indent\indent{\bf more in the following sections}\\
In the following sections  I shall also consider
some other scenarios that could fit my proposal of linking the OPERA anomaly
to issues related with evanescence/absorption and postselection.

In particular, I have so far neglected the role of neutrino oscillations, but
there may well be an interplay between absorption, postselection and neutrino oscillations.
Neutrino oscillations are themselves a cause of postselection, since they typically
produce situations such that only a fraction of the emitted particles can contribute
to the travel-time determination. And in this respect a particularly virulent role for
postselection would be present in the case of active/sterile neutrino oscillations.

There is even more room for postselection in OPERA-type setups. One could for example
shift the focus from the fact that the detected neutrinos must necessarily be
among those which were not absorbed in rock (at least not in the proximity of the detector)
to the fact that the determination of the travel time is performed only using
those particles that do interact with (or in the proximity of) the OPERA detector:
could that be seen as a source of tangible postselection?

I also consider the possibility that absorption and postselection could
act in synergy with dispersion. The propagation of neutrinos in
rock is of course characterized by both absorption and dispersion. And there are
several known experimental setups where
a superluminal ``speed of the peak of the distribution"
is produced by exploiting some aspects of the interplay between absorption and
dispersion~\cite{anomdispPRL,anomdispNATURE}.

After all these speculations, which may or may not apply to the understanding
of the OPERA anomaly, I argue that some of the topics here discussed must necessarily 
be relevant for a proper description of the
Planck-scale realm.

\section{More on travel times}
One of the main ingredients for the proposal I put forward in the previous section
is the description of time and particularly travel time in setups such that
there is significant interplay between the conceptual structure of quantum mechanics
and the conceptual structure of special relativity.

Time is notoriously troublesome in quantum mechanics. And
the part of relativistic quantum field theory which has been fully mastered
is the one concerning questions for the S-matrix, whereas most spacetime aspects
of relativistic quantum field theory remain the subject of endless controversy.
Example of this are studies~\cite{decaytime,decaytimetesta} of the short-time behaviour of an
unstable state weakly coupled to the final decay channels
and attempts~\cite{fermicausalONE,fermicausal,fermicausaltesta}
to verify relativistic causality in {\it gedanken} experiments
concerning the influence of the decay of a given excited atom
on another atom a certain distance away from it.

The main role of time in the OPERA setup is played by the ``travel time of the neutrinos".
And the notion of travel time is a particularly insidious aspect of time in quantum mechanics,
which often defies our intuition (inevitably bound to classical-limit experience).
Quantum mechanics (and therefore Nature, when properly understood) is about stages of
preparation and observation. It is not about anything ``going on in between preparation
and observation, while we are not looking":
as far as quantum mechanics is concerned, if nobody looked
then there is nothing to discuss about. We were thought that loud and clear with
double-slit and similar experiments. But it may be the case that we are still in the process
of fully appreciating how this affects the notion of travel time.\\
The notion of ``travel time of a particle" requires us to refer, at least implicitly,
to ``what is going on when nobody is looking": indeed it requires conceptualizing the particle
which is detected after traveling as ``the same particle" that was prepared at the emitter,
with an implicit (but strongly characteristic) assumption that one could follow the particle
as it travels.

While for free propagation such concerns may appear (and it is only appearance)
as merely academic, they take a more tangible aspect in other cases. Particularly clear
in this respect is the case of quantum tunneling, which has played the role of guiding
intuition for several of the points made in this manuscript.
Let us consider a quantum-mechanical barrier and let us prepare a pulse containing
very many electrons: only few electrons will reach a detector on the other side
of the barrier, and one is tempted to ask how much time it took them to travel
from the emitter to the detector through the barrier.\\
But of course there cannot be any scientific answer to such a question:
the electrons that are found on the other side
of the barrier cannot be viewed as some specific ones among those prepared.\\
Even in the case when we prepare a single electron and we end up finding ourselves in one of the
rare instances when then an electron is observed on the other side of the barrier it is still
not possible to argue that the one detected after the barrier is ``the same electron" prepared
before the barrier: electron-positron pair production plays a crucial role
(see, {\it e.g.}, Ref.~\cite{kleinparadoSingleElectron}) in the
relevant tunneling process. The only travel time
hosted by quantum mechanics which admits a classical-limit interpretation
is the signal/information travel time, but in general this comes without the
luxury of a classical-limit description of what carries the signal.

Evidently some travel times which are well defined quantum mechanically do not
have a corresponding classical-mechanics version.
Again the case of the tunneling times is particularly clear in this respect.
The (naive) question ``how much time it takes a particle to tunnel through a barrier?"
has generated confusion and debate since the early days of quantum
mechanics~\cite{earlytunnel1,earlytunnel2,bohmbook},
and it has attracted increasing interest in recent years, mostly because
of experimental studies where measurements of some travel times (but not the signal travel time)
have been found~\cite{stein2007,physrepREVIEW,recamiREVIEW,steinbergPRL,abruptNATURE}
to suggest superluminal propagation.
The tunneling time does not have a classical limit since classically there is no tunneling,
and if one formally insists on adopting classical notions in the analysis
a breakdown of applicability manifests itself with the fact that
the  momentum of the particle inside the barrier is imaginary.

This is particularly clear in the nonrelativistic limit.
Classically the time it takes a nonrelativistic particle to reach the other side of
a barrier (to go over a barrier) depends on the energy (and mass) of the particle
and the height of the barrier, since the kinetic energy is $E-V$ (nonrelativistic energy
minus barrier height).
For constant height of the barrier the time needed classically is
$$\frac{m}{p}L \simeq \frac{L}{\sqrt{2(E-V)}}$$
where $L$ is the length of the barrier.\\
But in the case of interest for quantum tunneling $E-V$ is negative
and the above expression becomes meaningless.

Attempts to introduce quantum mechanically a notion of travel time through the barrier
which at least to some extent could resemble our classical-limit perception of
the properties of travel time have produced a long list of candidates.
In addition to the ``barrier dwell time" and ``travel time of the peak of the distribution",
which I considered above, there is an impressive proliferation
of candidate characterizations of the ``tunneling
time"~\cite{stein2007,physrepREVIEW,recamiREVIEW,steinbergPRL,abruptNATURE},
some of which are those based on the use
of Feynman path integrals, those based on the Wigner distribution paths,
and those based on Larmor clocks.

The notion that most frequently applies to the time intervals 
usually measured experimentally
is the one I assume applies to the OPERA setup, {\it i.e.}
the ``pulse-peak travel time", which is the time scale obtained comparing
the timing  of the spatial centroid of
the initial/prepared distribution
and the timing  of the spatial centroid of
the final/measured distribution.
As mentioned on several occasions in this manuscript,
it is at this point well
established~\cite{stein2007,physrepREVIEW,recamiREVIEW,steinbergPRL,abruptNATURE}
that even this pulse-peak travel time does not in general give us
the travel time of information.
The aspect that misleads our classical-limit particle-propagation intuition
originates form the fact that
for smooth, frequency-band limited, distributions
the precursor tail of the distribution
allows to infer by analytic continuation the structure of the peak:
under such conditions, even for
free propagation, by the time the peak reaches a detector it carries
 no ``new information"~\cite{stein2007}
 with respect to the information already received through (hypothetical)
 detection of the precursor tail.
 It is indeed at this point also well
 established~\cite{stein2007,physrepREVIEW,recamiREVIEW,steinbergPRL,abruptNATURE},
 both experimentally and in theoretical pictures which are perfectly Poincar\'e invariant,
that the speed of the pulse peak (the speed derived from the pulse-peak travel time)
can be superluminal.
However when ``new information" is coded in
modified pulses containing ``abrupt" signals~\cite{stein2007,abruptNATURE},
experimentally one finds that these new-information features, if sufficiently
sharp, never propagate superluminally~\cite{abruptNATURE}.

It is perhaps characteristic of the confusion these issues generate
that the important and well measured
(though not giving the signal travel time) ``peak-pulse travel time"
is often called  ``phase time" in the literature,
which is very unfortunate since in experimental contexts where the comparison
is possible the velocity of the pulse peak does not reproduce the phase velocity,
but rather it reproduces the group velocity.

\section{Challenges for computing OPERA's ``peak travel time" from quantum field theory}
My proposal of analyzing the OPERA result from the viewpoint of an example of travel-time
issue (as for the tunneling time and other known peculiar travel-time effects) and particularly in relation to
evanescence and postselection, remains in this manuscript at a poorly formalized level.\\
In principle it does not take much to upgrade the proposal to a much more quantitative
status: it would suffice to compute the ``travel time of the peak" for the 
OPERA setup. 
But it seems to me that a reliable such
calculation can be very challenging because it should be done within
relativistic quantum field theory, with its notorious hostility toward providing crisp
spacetime characterizations. And probably one would have to take into account some of the
many-body aspects of the OPERA setup (many neutrinos in each bunch sent out from CERN,
many target nucleons along the way..).

Perhaps this can be done with a smaller effort than I presently foresee.
I have of course efforts in that direction ongoing,
and hopefully this manuscript will motivate others to try.
It appeared however appropriate not to postpone publicizing
my tentative picture of the OPERA anomaly because,
as stressed in the opening section,
it can be interestingly investigated experimentally, even without
a fully worked out model. And therefore even in the present form my speculations
could perhaps contribute to the ongoing debate on ``how to test/verify/falsify OPERA".

In this section I briefly note down some ideas for performing the actual computation
which is missing or at least some indicative approximation schemes.

\subsection{A generalized Fermi-causality paradox?}
The derivation I wish I was able to offer in this manuscript is the one
of  the ``travel time of the peak" for the OPERA setup, formalized within
relativistic quantum field theory.

It is intriguing to notice that in any case (whether or not the type of effect I here
propose is actually present) the OPERA setup represents a generalization (within a {\underline{real}}
experiment) of a type of application of relativistic quantum field theory
which was originally contemplated by Fermi within a {\it gedanken} experiment.\\
Fermi studied~\cite{fermicausality} the
effects induced by the decay of an excited atom A
on another atom B placed at a certain distance R.
This was indeed Fermi's way to test relativistic causality within quantum field theory.
Within the approximation adopted by Fermi one does find that no influence
can be transmitted from A to B before a time
R/c has elapsed.

The situation is less clear if one goes beyond the level of description
afforded by Fermi's approximations: one then finds that
apparent violations of causality are present~\cite{fermicausalONE,fermicausal}
even though at a more careful level of analysis one finds~\cite{fermicausaltesta}
that rather than violations of causality such computations expose
 the lack of notion of sharp localizability for a relativistic quantum system.

 The type of analysis that is needed for my purposes is essentially
 analogous conceptually to studies of causality in Fermi's {\it gedanken} setup,
 but is much more complex: Fermi's setup involves only two atoms, an emitter atom
 and a detector atom in vacuum, whereas investigating my thesis for OPERA would require
analyzing with comparable rigor the case of the CERN emitter and the LNGS detector
exchanging neutrino signals in rock. Considering how Fermi's {\it gedanken} setup
was hotly debated, in spite of its apparent simplicity, it seems natural to
assume that consensus on causality for the OPERA setup will not materialize quickly.

\subsection{How much time does it take to travel through an imaginary barrier?}
Of course, one may try to proceed
at first allowing some room for semi-heuristic arguments. A potentially amusing 
example of this would
be the study of the Dirac equation for the description of a pulse of particles
encountering an ``imaginary barrier": heuristically one could take as starting
point the techniques of analysis that are familiar for the case
of the study of the Dirac equation in presence of a real potential barrier
(see, {\it e.g.}, Refs.~\cite{diractunnel1,diractunnel2})
but replace the real potential barrier $V$ with a purely 
imaginary potential ``barrier" ${\tilde V}$.
And the imaginary barrier could be made energy dependent, in order to model the energy
dependence of absorption in rock of OPERA neutrinos.\\
The quantity of interest in such an analysis would be of course the ``travel time
of the pulse peak" in the sense discussed above.

With such a heuristic setup one would attempt to mimic the effects of rock
on neutrino propagation, within the assumption that absorption is the feature
of primary interest in the OPERA case. It may well be too crude a model of the actual
derivation which is needed for my purposes, but it is nonetheless an intriguing 
exercise of relativistic quantum mechanics, using the Dirac equation (as if particle production
was negligible) and yet allowing for absorption of part of the initial pulse.

\section{Other aspects of the phenomenology}
In the opening section I stressed how, in spite of its preliminary status of development,
my proposal has potentially intriguing implications for the phenomenology of OPERA-like
setups with neutrinos and antineutrinos.
But, as already shown by these first months of discussion of the OPERA anomaly,
any proposed ``solution" of the OPERA anomaly must also be successful in addressing
potential challenges from other known experimental facts.

In this brief section I mainly comment on the challenges of this sort which have been
most discussed in the OPERA literature.

\subsection{No large effects for SN1987a neutrinos}
For most solutions of the OPERA anomaly which have been attempted in the recent literature
a severe challenge is provided~\cite{whataboutopera,giudice,operaELLIS}
 by the fact that the observations of supernova SN1987a
suggest rather stringent bounds on superluminality of 
neutrinos.

Agreement with observations of SN1987a is instead assured for the type
of mechanism I am advocating since very little absorption/postselection 
would be relevant for the
analysis of neutrinos observed from distant supernovas.

\subsection{No anomalies for pion decay}
For description of the OPERA result based on superluminal neutrinos breaking Lorentz symmetry
a key challenge is found in the analysis of the
 pion-decay process $\pi \rightarrow \mu + \nu$ in vacuum: one would find that
 for pions of sufficiently high energies
  the  muon/muon-neutrino phase space
available for the decay is severely reduced~\cite{gonzaOPERA,bietal,cowsiketal}.

It has been established that alternative scenarios
in which superluminal neutrinos are introduced deforming Lorentz symmetry
are automatically immune from this challenge~\cite{operaDSR}.

Also the scenario I have here proposed is immune from this challenge, since no tangible role
for absorption and postselection could be present in such pion-decay processes.

\subsection{Probably no Cherenkov-like processes}
Another challenge to 
descriptions of the OPERA result based on superluminal neutrinos breaking Lorentz symmetry
 is found in the possibility of Cherenkov-like processes in vacuum,
 such as $\nu \rightarrow \nu + e^+ + e^-$.
 When the neutrino is superluminal and the framework is such that Lorentz symmetry
 is broken such processes become allowed above a certain threshold energy~\cite{cohenglashow}.

Also this challenge is automatically evaded by scenarios
in which superluminal neutrinos are introduced deforming Lorentz symmetry~\cite{operaDSR}.

For what concerns the proposal I am here putting forward it is evident
that processes like $\nu \rightarrow \nu + e^+ + e^-$ in vacuum would still not be allowed
(and therefore represent no challenge) since no tangible role
for absorption and postselection could be present.
For the case of a scenario such as the one I am here
advocating it is more interesting to consider 
the possibility of $\nu \rightarrow \nu + e^+ + e^-$ in rock.
I postpone a more careful analysis of this possibility to a future study.
I here notice that for such processes in rock one could not exclude a priori
a role for absorption and postselection, but I expect that the process remains forbidden.
In the picture I am advocating the ingredients used in Ref.~\cite{cohenglashow} for the 
onset of $\nu \rightarrow \nu + e^+ + e^-$ should be missing, even in rock, since
the apparent superluminality reported by OPERA would be only a manifestation of
the postselection performed on the initial neutrino pulse:
I do not expect any particle to be transferred to a non-special-relativistic shell,
and I do not expect any particle to actually ``travel superluminally".

\subsection{More on testing the postselected-velocity subluminality bound}
I have already mentioned in the opening section some of the possible ways to test
with neutrinos and antineutrinos
my conjectured subluminality bound for the postselected velocity.
This conjecture however is applicable to a variety of contexts, whether or not the OPERA
anomaly is described in terms of absorption and postselection.
Actually the most natural contexts to probe this conjecture would be cases where
a first-quantized description is applicable: my conjecture is 
$$\mathsf{v}_{corr} = {\cal P}_{select} \mathsf{v}_{select} \leq 1$$
and it evidently only makes sense if the experimental setup is such that 
one prepares a certain number of particles (using classical-limit language)
and some of these initial particles contribute to the velocity (of the pulse peak) measurement.\\
Any experimental context with mechanisms of ``gain" of the signal would of course 
go beyond the possible realm of applicability of the conjecture.

But evidently even in cases where a first-quantized description is legitimate the
validity of the conjecture is in no way assured, and could be scrutinized both experimentally
and at the theory level. To see this very clearly let us again consider an idealized situation
of quantum-tunneling type: a situation where an initial pulse is prepared and
there are two widely separated pulses observed ({\it e.g.} the transmitted pulse and the
reflected pulse, for quantum tunneling).\\
Then one could assume a description of the average velocity for the entire ensemble
roughly given by
$$<\mathsf{v}> = P_1 <\mathsf{v}_1> + P_2 <\mathsf{v}_2>$$
with indices $1$ and $2$ for the two outgoing pulses.\\
In such an idealized case my conjecture would amount to finding 
that both $P_1 <\mathsf{v}_1> \, \leq \, 1$ and $P_2 <\mathsf{v}_2> \, \leq \, 1$. 
However, if truly we should
only enforce $<\mathsf{v}> \, \leq \, 1$ then, say, one could still
have $P_2 <\mathsf{v}_2> \, = \, 1 + \Delta$
(with positive $\Delta$, therefore a violation of my conjecture) 
if it happens to be the case
that correspondingly $P_1 <\mathsf{v}_1> \, \leq \, 1 - \Delta$.\\
So I expect that my conjecture is not protected by any general principle, 
but it appears to be verified
in all cases that I know of with superluminality (of the pulse-peak velocity) 
without gain. I hope some readers
will be interested in testing this.

\section{Possible variants for involving evanescence/absorption and postselection}\label{othertunnels}

\subsection{More detailed aspects of absorption}
In the simplest version of the scenario I am proposing
 a key role would be played in the OPERA result by the absorption of
neutrinos in rock and its implications from a postselection perspective.
The tentative estimates I gave in the opening section all focused on the $\xi$
that can be computed as a characterization of the
overall amount of neutrino absorption present in a given OPERA-like setup.
But absorption and postselection could play a more subtle role, with a central role
being played by the energy dependence of the neutrino cross section.\\
From this perspective it is interesting to notice that 
most of the energy range over which the
OPERA data are collected is in a region where the neutrino cross section
has a rather complicated energy dependence~\cite{liparilusignoli,pdg}.
For energies of $\sim 50GeV$ and above the dependence on energy of the cross section
turns to a very simple law, very close to linear dependence.\\
It is therefore plausible, if in particular derivatives with respect to energy
are relevant, that there would be significant differences between studies involving
 neutrinos with energies between $1 GeV$ and $50 GeV$ and analogous studies
 with energy range between, say, $50 GeV$ and $100 GeV$, differences even bigger than the
 ones that could be already expected on the basis of the difference
 in overall absorption. 

\subsection{Interplay between dispersion and absorption}
I feel one should also consider the possibility that 
in OPERA-like setups absorption and postselection could
act in synergy with dispersion. The propagation of neutrinos in
rock is of course characterized by both absorption and dispersion. And there are
several known experimental setups where
a superluminal ``speed of the peak of the distribution"
is produced by exploiting some aspects of the interplay between absorption and
dispersion~\cite{anomdispPRL,anomdispNATURE}.

Actually in Ref.~\cite{brusteinOPERA} it was argued that the effect
seen by OPERA might be due to dispersion\footnote{The argument in Ref.~\cite{brusteinOPERA}
relied rather crucially on an estimate based on the number of neutrinos in each of the bunches contained
in every extraction used for the first OPERA measurement (reported September 2011). It seems that
such an estimate no longer holds for the second OPERA measurement (November 2011), since then each bunch
contained fewer neutrinos.} in rock, ``amplified" by the fact that
the neutrinos prepared for OPERA studies
are in a coherent quantum state with a large intensity.

Since it is established that some striking examples of superluminality of the ``pulse-peak velocity"
can be achieved exploiting the interplay between absorption and dispersion,
it could be interesting to combine some of the elements of the thesis reported
in Ref.~\cite{brusteinOPERA} with my thesis for absorption and postselection.

\subsection{Postselection with neutrino oscillations}
I have so far ignored an aspect of neutrino physics which could complicate matters rather significantly
from the perspective of my proposal: neutrino oscillations.\\
These oscillations introduce themselves some opportunities for
reasoning on OPERA from  a postselection perspective: a pulse of neutrinos is prepared which
mixes different types of neutrinos, but then the detector is structured so that only one type
of neutrinos can be revealed.\\
If there were active/sterile neutrino oscillations then this sort of postselection
would be particularly evident, since the detector cannot reveal the sterile ones.

This opens the possibility of several generalizations of the scenario I have here adopted.
I do not see how the postselection due to neutrino oscillations could by itself
produce the effect seen by OPERA, but this should be investigated quantitatively.
Perhaps one could contemplate some role for an energy dependence of the oscillations.\\
And in any case there could be an interplay between the sort of postselection that results 
from absorption in rock and the sort of postselection that results from oscillations.

\subsection{On the role of detector-postselection}
One more possibility of postselection mechanism I should mention 
concerns the detector itself.\\
In the scenario I proposed in the opening section the postselection mechanism
is centered on the fact that OPERA's ``velocity" measurement is not performed
on all neutrinos, but only on those that do manage to avoid absorption in rock.\\
An alternative is to consider the selection at detection: it is also true that
the OPERA result only uses data on those neutrinos that interact with the detector
(or with rock in proximity of the detector). 
These are only very few neutrinos within the original pulse.
A situation much different from the type of selection that inevitably also other
particle detectors perform, since for other particles the efficiency of detection
is always much higher than in the neutrino case.

Whether or not it ends up being relevant for understanding the OPERA anomaly,
I believe that the study of OPERA-like setups for what concerns these aspects
of postselection is of significant interest, even just from the theoretical perspective,
for the understanding of a special case of exchange of signals from an emitter
to a detector. The most interesting aspect from this perspective comes from the combination
of two features: the ``survival probability" (probability of avoiding absorption in rock
while ``traveling" from CERN to LNGS) depends on energy, and also the ``detection probability"
(probability of being stopped/revealed at the LNGS detector or in rock in its proxomity)
depends on energy. 
Because of the properties of the neutrino cross section in that energy 
range~\cite{liparilusignoli,pdg},
the survival probability decrease with energy while the detection probability
increases with energy.

\section{Relevance for the Planck-scale realm}
I have here speculated that some aspects of the interplay between special relativity and quantum
mechanics, some of those most subtle from the information-theory viewpoint,
might be responsible for the OPERA anomaly. I feel this needs to be checked more carefully
than done so far in the literature, but of course it is just one of the paths
that could lead to an understanding of the OPERA result (among which, of course,
one should still consider as most likely the possibility of an unnoticed source of
systematic error).

Something which I feel we can be more certain about is the relevance of the
type of issues and observations I here discussed for the understanding of the Planck-scale
realm.
Evidently these speculations which I have here reported are not related in any way
(and not inspired by) aspects of the quantum-gravity problem. On the contrary,
as I am working on this ongoing project (of which I here reported a few preliminary observations)
I find that it should be the quantum-gravity side, those working on candidate solutions
of the quantum-gravity problem, that should take notice of the issues and observations
which were the main focus of this manuscript and of some of the references here cited.

Primarily this expectation originates from a perception of how 
the consistency of the present logical picture of the ``laws of fundamental physics" 
is ensured only by a
very delicate 
balance among the ingredients used in the formalization of special-relativity and quantum mechanics.
The fact that we expect for measurements with Planck-scale accuracy (presently
only hypothetical) to manifest a new ``fuzzy" picture of spacetime appears to
provide a strong invitation for reassessing whether this delicate balance
of formal ingredients could survive the transition to 
a Planck-scale-accurate spacetime picture.

To mention explicitly at least one example let me consider the role played 
(explicitly or implicitly) in present analyses of relativistic (and quantum mechanical)
exchange of signals by the availability of two powerful abstractions
(see, {\it e.g.}, Ref.~\cite{steinREALITY}):
on one side we are allowed to abstract the possibility
of signals with infinite tails in
the frequency domain and arbitrarily abrupt changes in time, and on the other side we are allowed
to abstract signals with infinite ``precursors" in time and 
sharp cutoffs in frequency.
At least two of the most popular expectations for the Planck-scale realm appear
to necessarily affect the balance of such abstractions: the expectation that the
Planck scale should set the maximum allowed value for energy (and/or momentum) 
of fundamental particles,
and the expectation that the fundamental description of spacetime should
not afford us the luxury of attributing physical meaning to evolution
on time/spatial scales smaller than the Planck length.
It appears then that in the Planck-scale realm
we should renounce to the idealization of signals with 
infinite tails in
the frequency domain and arbitrarily abrupt changes in time.

And complementary concerns apply to the idealization of
signals with infinite ``precursors" in time and
sharp cutoffs in frequency.
 For that idealization I see as main Planck-scale challenge the fact that
 the infinite but faint precursors in time could not have the same significance
 in the ``fuzzy" Planck-scale realm that they instead have in the present
 sharp picture of spacetime.
 To see this most vividly I can resort to one of the observations used in this manuscript:
 quantum tunneling may result in the fact that the information usually (for free propagation)
 coded in the precursor tail of a peaked pulse is instead reshuffled in an ``early arrival"
 peaked pulse. If we can model in such instances the fuzzyness of spacetime as a source of
 noise~\cite{gacnoise,ngnoise} then these two ways of carrying the same amount of information 
may loose their ``information balance": in presence of an irreducible noise source
 the information coded in a long but faint 
 precursor tail cannot be viewed as information actually reaching a distant detector,
 whereas the reorganization of the same information in the form of the ``early arrival"
of a peaked pulse could be less subject to the implications of the noise effectively
introduced by spacetime fuzziness.

So I feel~\cite{gacinprep} that a more vigorous effort should be directed toward understanding
how gravitational effects at the Planck scale can change or affect the description
of relativistic quantum mechanics as a theory of relativistic propagation
of ``waves of information". There are several early signs that this might gradually 
take center stage, among which I mention as examples 
studies of black holes form the viewpoint of quantum information
(see, {\it e.g.}, Ref.~\cite{jackElloyd} and references therein),
and Kemp's proposal of a dual description of spacetime, 
at once continuous and discrete~\cite{kempfshannon}.
Particularly the approach of Ref.~\cite{kempfshannon}
shares some of the intuition I have here adopted (within an otherwise completely
unrelated proposal). The main observation of Ref.~\cite{kempfshannon}
is based on Shannon's sampling theorem~\cite{shannon},
showing that a bandlimited signal can be perfectly reconstructed everywhere
by recording only some discrete samples of it,
and the observation that
physical fields on generic curved spaces obey a sampling theorem if they possess an
ultraviolet cutoff.

\section{Schematic summary}
I have here adopted a rather speculative attitude, and I took the liberty 
of reporting what is only a preliminary
exploration of a scenario for interpreting the OPERA velocity measurement. 
I feel this is justifiable in light of the potential significance, in any case,
of the OPERA result  (if correct...of course we should still assume as most likely
the possibility that some unnoticed systematic error be responsible for the reported data).
In closing, I offer a sketchy summary of
the main points here made.

\smallskip

\noindent
Concerning the assessment of the OPERA result from any perspective:\\
$\star$ The facts so far established in the related literature do not allows us to conclude
that the velocity measured by OPERA is a ``signal velocity". Superluminal measured
values for a signal velocity would challenge special relativity, but the velocity
measured by OPERA has (at least some of) the properties of a ``peak-pulse velocity",
and superluminal measured
values for a peak-pulse velocity can be compatible with ordinarily special-relativistic causality.\\
$\star$ The facts so far established in the related literature do not allows us to conclude
that neutrino absorption in rock has a negligible impact on the OPERA velocity measurement.
For OPERA neutrinos absorption in rock is an effect amounting to a few parts in $10^5$
and therefore its smallness is comparable to the smallness of the ``velocity anomaly" found
in the OPERA results.\\
$\star$ In turn, especially since absorption cannot be automatically neglected,
the facts so far established in the related literature do not allows us to exclude a role
in the analysis of the OPERA result of some of the concepts known for the study
of postselected/``weak" quantum measurements. The OPERA measurement appears to be not
really ``weak" but also not fully ``strong" (the notion of ``quasi-strong measurement" which
I described in the opening section).

\smallskip

\noindent
Concerning the description of the OPERA result which was here explored:\\
$\star$ I have proposed that the findings reported by OPERA could be described
in terms of travel-time features connected with ``evanescence" and postselection,
similar to (but ultimately different from) features found in studies of ``superluminal
tunneling times".\\
$\star$ Even for the crudest hypothesis, according to which absorption in rock would
be primarily responsible for the effect, I am still unable to compute the
needed ``pulse-peak travel time" for an OPERA-type setup. I only managed to provide
a tentative estimation of the ``rock dwell time", and this did turn out to be in rough
agreement with the travel times reported by OPERA. This may be encouraging since
for the study of quantum tunneling in several (though not all)
 possible setups the ``barrier dwell time"
is in  rough agreement with the ``pulse-peak travel time through the barrier".\\
$\star$ According to this crudest formulation of my scenario experimental
results for OPERA-like setups involving antineutrinos should find effects
that are comparable to the ones for neutrino but quantitatively tangibly different
from the ones for neutrinos. And in comparing different OPERA-like studies of
neutrinos one would expect, especially for what concerns the dependence on the energy
of the neutrinos and the dependence on the emitter-detector distance,
some characteristic qualitative features, much different from the ones expected
if the OPERA result is interpreted as a determination of a ``physical particle velocity"
(in the standard sense borrowed from classical-limit intuitive pictures).\\
$\star$ It appears likely that absorption in rock and postselection would 
interface with some other of the many features of the OPERA measurement, which is
a measurement of significant complexity if viewed from the perspective of the
combined implications of special relativity and quantum mechanics.
This opens the possibility for several more refined scenarios, some of which
have been contemplated briefly in Section~\ref{othertunnels}, attributing to 
absorption in rock and/or postselection only a ``partial responsibility" for the OPERA anomaly.

\section*{ACKNOWLEDGEMENTS}
\noindent It was a
%an invaluable 
privilege
to discuss some of the results reported in
Refs.~\cite{liparilusignoli,decaytimetesta,fermicausaltesta}
with Paolo Lipari and Massimo Testa

\end{document}